\documentclass[prl,a4paper,superscriptaddress,reprint]{revtex4-1}  
\usepackage{graphicx}
\usepackage[english]{babel}
\usepackage{amsfonts,amssymb,amsmath}
\usepackage[dvipsnames]{xcolor}

\begin{document}

\title{Staging superstructures in high-$T_c$ Sr/O co-doped La$_{2-x}$Sr$_x$CuO$_{4+y}$}
\date{\today}

\author{P. J. \surname{Ray}}
\email[Corresponding author: ]{pia@pjray.dk}
\affiliation{Nanoscience Center, Niels Bohr Institute, University of Copenhagen, DK-2100 Copenhagen, Denmark}

\author{N. H. Andersen}
\affiliation{Physics Department, Technical University of Denmark, DK-2800 Kgs. Lyngby, Denmark}

\author{T. B. S. Jensen}
\affiliation{Physics Department, Technical University of Denmark, DK-2800 Kgs. Lyngby, Denmark}

\author{H. E. Mohottala}
\affiliation{Department of Physics, University of Connecticut, U-3046, 2152 Hillside Road, Storrs, Connecticut 06269-3046, USA}
\affiliation{Department of Physics, University of Hartford, 200, Bloomfield Ave., West Hartford, CT-06117, USA}

\author{Ch. Niedermayer}
\affiliation{Laboratory for Neutron Scattering, ETHZ \& PSI, Ch-5232 Villigen PSI, Switzerland}

\author{K. Lefmann}
\affiliation{Nanoscience Center, Niels Bohr Institute, University of Copenhagen, DK-2100 Copenhagen, Denmark}

\author{B. O. Wells}
\affiliation{Department of Physics, University of Connecticut, U-3046, 2152 Hillside Road, Storrs, Connecticut 06269-3046, USA}

\author{M. v. Zimmermann}
\affiliation{Deutsches Elektronen-Synchrotron DESY, Notkestr. 85, 22603 Hamburg, Germany}

\author{L. Udby}
\affiliation{Nanoscience Center, Niels Bohr Institute, University of Copenhagen, DK-2100 Copenhagen, Denmark}

\begin{abstract}
  We present high energy X-ray diffraction studies on the structural phases of an optimal high-$T_c$ superconductor La$_{2-x}$Sr$_x$CuO$_{4+y}$ tailored by co-hole-doping. This is specifically done by varying the content of two very different chemical species, Sr and O, respectively, in order to study the influence of each. 
  A superstructure known as staging is observed in all samples, with the staging number $n$ increasing for higher Sr dopings $x$.   
  We find that the staging phases emerge abruptly with temperature, and can be described as a second order phase transition with transition temperatures slightly depending on the Sr doping. 
  The Sr appears to correlate the interstitial oxygen in a way that stabilises the reproducibility of the staging phase both in terms of staging period and volume fraction in a specific sample. 
  The structural details as investigated in this letter appear to have no direct bearing on the electronic phase separation previously observed in the same samples. This provides new evidence that the electronic phase separation is determined by the overall hole concentration rather than specific Sr/O content and concommittant structural details. 
\end{abstract}

\maketitle


The rich phase diagram of the superconducting cuprates continues to motivate new investigations for the mechanism behind high-$T_c$ superconductivity in these materials~\cite{Ghiringhelli2012,Chang2012,Torchinsky2013,Comin2014,SilvaNeto2014,Thampy2014,Keimer2015,Kremer1993}. 
In samples where hole-doping is performed by chemical substitution of La for Sr to give La$_{2-x}$Sr$_x$CuO$_4$, it is well known that the details of the superconducting and magnetic phases strongly depend on $x$.
Evidence for electronic phase-separation, which can possibly be explained by the formation of percolative hole-networks~\cite{Kremer1992,Kremer1994}, has been given in samples which are instead hole-doped with oxygen to give La$_2$CuO$_{4+y}$. The details depend on $y$, but one electronic phase is a bulk superconductor with discrete $T_c$ in the range 30-45 K, while the other is an antiferromagnetic phase which is commensurate for $y < 0.055$ and modulated with $\delta\sim 0.10$-$0.13$ for higher dopings~\cite{Khaykovich2002,Khaykovich2003,Kremer1994,Wells1997}.
A range of superstructures measurable by diffraction techniques form as a consequence of oxygen intercalation. In highly oxygenated La$_2$CuO$_{4+y}$ the details and spatial distribution of the superstructures as well as $T_c$~\cite{Fratini2010,Poccia2012} and the strength of the modulated antiferromagnetic signal~\cite{Lee2004} depends strongly on the cooling rate. It would thus be practical to study other systems which are electronically similar when at low temperature equilibrium, but which can be reached at natural cooling rates. This is a major motivation for the present study.

\begin{figure}[tbp]
  \includegraphics[scale=1]{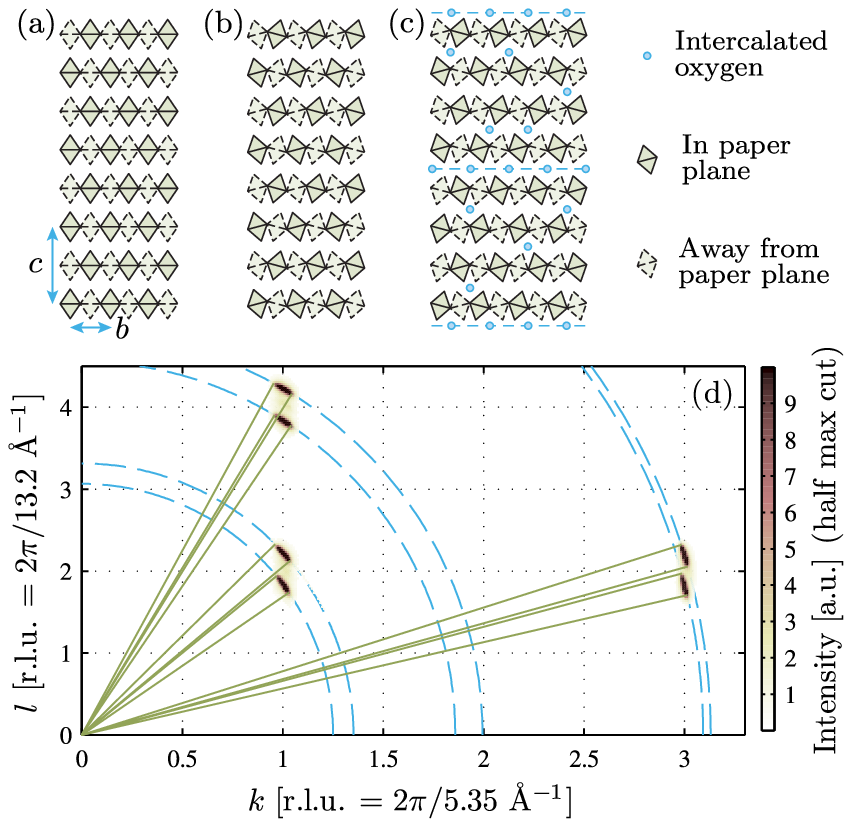}
  \caption{Illustration of relevant structures, with only CuO$_6$ octahedra and interstitial oxygen shown. 
  (a)~Tetragonal \emph{F4/mmm} and 
  (b)~ortorhombic \emph{Bmab} structures of La$_{2-x}$Sr$_x$CuO$_4$ seen along the $a$ axis.  
  (c)~Example of a staged structure caused by layers of favourable positions for interstitial oxygen due to tilt flip domain borders. Figure inspired by~\cite{Wells1997,Wells1996,Tranquada1994}. 
  (d)~Map of investigations of the $[0KL]$ plane for the $x=0.04$ sample at $T=10$~K, close to \emph{Bmab} allowed positions (012), (014), and (032). The intensity of each pair of peaks has been scaled for visibility. 
  Blue dashed arcs are centered in origo, showing how the resolution function is widest transverse to $q$. 
  Green lines mark out the FWHM of the long axis for each peak. }
  \label{fig:stagingtilts}  
\end{figure}

In setting the stage for the work in this letter, we recap important findings in the isostructural compound La$_2$NiO$_{4+y}$, where the intercalated oxygen density tends to modulate along the $c$ axis in a superstructure known as staging~\cite{Tranquada1994}. The superstructure is characterised by a number $n$ referring to the periodicity of the intercalated oxygen layers in terms of NiO$_6$ spacings. A similar superstructure has been observed in La$_2$CuO$_{4+y}$~\cite{Wells1996,Lee1999}, and in both compounds the staging number is expected to decrease with increased oxygen concentration along with smaller shifts associated with increased quenching temperature~\cite{Lee2004,Wells1996,Lorenzo1995}.
An illustration of this superstructure is shown in Fig.~\ref{fig:stagingtilts}(a-c). 

In order to study the role of the mobile versus localised dopants we have prepared single crystals of La$_{2-x}$Sr$_x$CuO$_{4+y}$, hole-doped to a level of $0.125 \leq n_h \leq 0.16$ by a combination of Sr chemical subsitution and intercalation of O~\cite{Mohottala2006}. We have previously investigated the superconducting and magnetic properties and found that a unique superconducting phase and a unique stripe-like magnetic phase is present in all Sr/O co-doped samples with remarkably coinciding transition temperatures $T_c \approx T_N \approx 40$~K regardless of $x$~\cite{Mohottala2006,Udby2013PRL}. 
See {supplementary material}~\cite{SupplementalMaterial} for an estimate of the oxygen contents. 

In this letter we investigate the structural impacts of this co-hole-doping. 
We find that each of the investigated samples exhibit staging, or hints of staging.


The samples studied in this letter are the same single crystals with Sr doping $x=0.00$, 0.04, 0.065, and 0.09 as first studied in Ref.~\cite{Mohottala2006}. The $x=0.00$ crystal was grown in a crucible, while the others were grown by the travelling solvent floating zone method. 
The crystals were electrochemically oxidised through electrolysis in a NaOH solution bath~\cite{Wells1996}. 

X-ray scattering data presented in this letter were performed at the high energy X-ray 
triple-axis diffractometer BW5~\cite{BouchardBW51998} at the now decommissioned DORIS III storage ring at DESY, Germany. 
The beam size for this instrument at the sample position was between 1$\times$1 and 2$\times$2~mm$^2$, while the used samples had typical dimensions of 1 to 5~mm. 
In all experiments the crystals were cooled with $<2$~K/min unless specifically stated otherwise. The slow cooling rate should allow reproducible ordering of the intercalated oxygen~\cite{Lee2004}. 

Throughout this letter the Miller indices refer to the orthorhombic \emph{Bmab} notation ($a<b<c$, $a$ and $b$ in the range 5.3-5.4~\AA{} and $c$ in the range 13.0-13.2~\AA{} for our samples), also 
above the transition temperature where the unit cell is described in the space group \emph{F4/mmm}.


We have mapped areas of the $(0KL)$ plane close to \emph{Bmab} allowed positions in a Sr/O co-doped crystal with $x=0.04$ by sample rotation scans, as shown in Fig.~\ref{fig:stagingtilts}(d). Superstructure peaks are observed with positions corresponding to a modulated structure along the $c$ axis in a similar manner as the staging in La$_2$NiO$_{4+y}$. The observed superstructure peaks are broadened transverse to the scattering vector -- along the dashed arcs -- with FWHM between $1.4$ and $3.4^\circ$, and specifically $2.25(5)^\circ$ for the staging peak pairs close to (014).
For comparison the transverse spread of the close-by fundamental Bragg reflection (004) is $\text{FWHM}=0.3^\circ$, which means the transverse broadening is intrinsic to the superstructures. Along the scattering vector of the superstructure peak the FWHM is between 0.25 and 0.34~\AA$^{-1}$, which is much more than the $\text{FWHM}=0.009$~\AA$^{-1}$ of close-by (004). 
This means that the distribution of unit cell lengths in the staged volume of the sample is broader than for the unstaged volume. This is also reflected in the non-integer values and distribution width of the staging period values discussed below.

\begin{figure}[tbp]
  \includegraphics[scale=1]{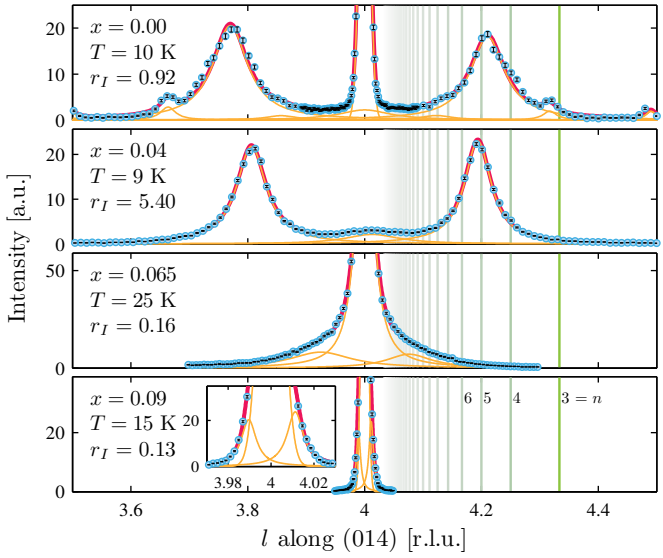}
  \caption{Low temperature data for the four samples. The plots for the $x=0.00$, 0.065, and 0.09 samples are cut off from the maximum intensities of 101.4, 196.5, and 196.9, respectively, to show the staging peaks. The inset plot in the bottom panel shows a zoom for clarity. Black lines indicate errorbars, and are in general smaller than the data symbols. Red thick curves show full fits of the spectra, composed of the sum of individual peaks (thin orange curves) and a flat background. Vertical green lines indicate integer $n$ staging. The relative intensity $r_I$ is calculated from the integrated intensities $I_\text{staging} / I_\text{\emph{Bmab}}$, with the former being the average of the two largest staging peaks. }
  \label{fig:lowtempoverview}
\end{figure}

In order to identify and characterise the superstructures as function of Sr content, further measurements were done by scans along $l$ through the \emph{Bmab} allowed reflection (014) in the four samples.
Fig.~\ref{fig:lowtempoverview} shows examples of these scans at low temperatures for each of the samples. 
In all samples a peak at the \emph{Bmab} allowed position (014) is observed. 
In order to analyse the data further, we relate the reciprocal distance from the (014) position, $\Delta l$, to the staging number $n$, by $n=1/\Delta l$. 

For the $x=0.00$ sample, several individual staging peaks were observed on both sides of the central \emph{Bmab} position. 
The largest peaks correspond to a staging number $n$ between 4 and 5, while smaller peaks were observed with $n$ just above 3 and between 7 and 8 on either side of this. 
For the $x=0.04$ sample, clear staging peaks with $n$ just above 5 were observed on either side of the anomalously small central \emph{Bmab} peak. 
For the $x=0.065$ sample, the high shoulders of the \emph{Bmab} peak could be fitted with staging peaks with $n$ between 13 and 14.   
Although no staging is visible to the naked eye in the data for the $x=0.09$ sample, we found that the best fit of the data was to a large central \emph{Bmab} peak with small shoulders reminiscent of staging with a very high $n$. 

\begin{figure}[tbp]
  \includegraphics[scale=1]{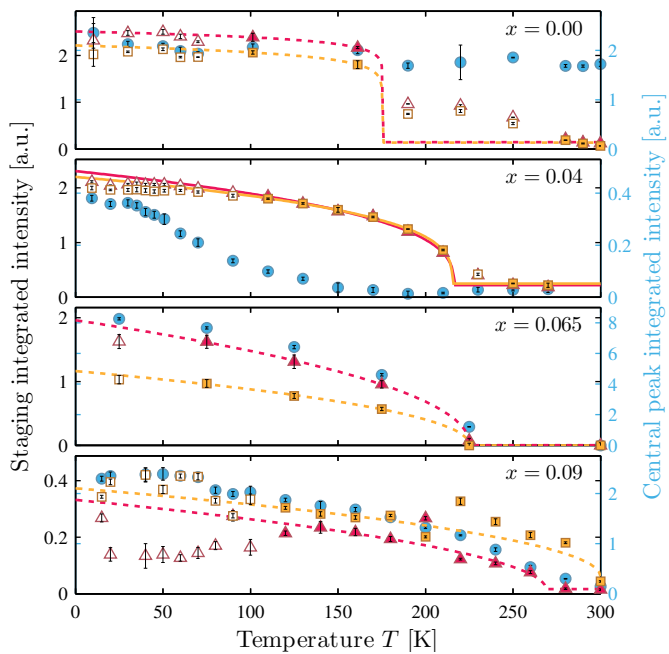}
  \caption{Temperature dependence of integrated intensities of staging peaks (red triangles for low $l$ and yellow squares for high $l$) and central peaks (blue circles). Lines for the staging data are power law fits to the filled markers only. Empty markers are excluded from the fits due to order parameter saturation (low temperatures) and critical scattering (close to transition temperatures).}
  \label{fig:tempdependencies}
\end{figure}

Scans similar to those in Fig.~\ref{fig:lowtempoverview} were taken at various temperatures and the data was fitted to the best Lorentzian and/or Gaussian peak shapes in order to obtain individual integrated intensities, positions, and widths for each peak. The integrated intensities for the central \emph{Bmab} peak and the staging peaks (for the $x=0.00$ sample only the most intense pair of staging peaks) are shown in Fig.~\ref{fig:tempdependencies}. 
Each dataset shows 
phase transitions with varying transition temperatures and decay rates for the \emph{Bmab} and staging peaks, respectively. Note, however, that the staging peaks are comparable in intensity and widths by pairs, and that for all samples the staging transition appears quite abruptly. The $x=0.00$ \emph{Bmab} peak data did not show a phase transition below room temperature, which is in accordance with earlier investigations~\cite{Radaelli1994}. 

The staging transition is most clearly interpreted in the $x=0.04$ sample, wherefore we have chosen to analyse these data in more detail and generalise the observations to the other samples where applicable. For each temperature the scan was fit to three Lorentzians (with positions and widths shown in the {supplementary}). 
A small difference observed between the position of the observed \emph{Bmab} peaks and the center between the positions of the staging peaks indicated a phase separation between the staged and non-staged parts of the samples with slightly different $c$ axis lengths. Furthermore, since the staging transition temperature is higher than the \emph{Bmab} transition temperature at least in the $x=0.04$ sample, staging cannot be a superstructure of \emph{Bmab} within the same sample volume. This confirms earlier results on oxygen-only doped compounds~\cite{Jorgensen1988,Wells1996,Lee2004}. As the relative intensity of the (014) and staging peaks seen in Fig.~\ref{fig:tempdependencies} does not depend monotonically on Sr content we also believe that the specific amount of staged/non-staged volume in each sample does not depend (monotonically) on Sr content. 
Finally, the width of the staging peaks were seen to diverge close to the transition temperature, as expected from a second order phase transition. 

The staging temperature data in Fig.~\ref{fig:tempdependencies} were fit with a power law as shown for the $x=0.04$ sample with a solid line. 
We use a method commonly used for perovskites~\cite{Hunnefeld1998} to exclude low temperature data points due to saturation of the signal, and data points just above the transition due to critical scattering. 
The power law fit yields $T_\text{staging}=215.8(7)$~K, which is a significantly smaller transition temperature than in the nickelates~\cite{Tranquada1994}. 
The obtained critical exponents $2\beta$ for the staging peaks are $0.312(14)$ and $0.36(2)$, respectively, which falls between 2D and 3D Ising models, leaning closer to the 2D model~\cite{BlundellMagnetism}. This leads us to speculate that similar 2D directional kinetics could be behind the formation of the staging superstructures. 

In order to analyse staging in the more difficult datasets for some of the other samples, it was necessary to use a gradient analysis method (detailed in the {supplementary})  
to fix the transition temperature before fitting with a power law, the result of which are shown as dashed lines in Fig.~\ref{fig:tempdependencies}. 
We observe an indication of slightly increasing transition temperature with $x$ and comparable critical exponents -- with exception of the improbably low exponent we found for the $x=0.00$ sample,  possibly caused by our exclusion of data points due to a second, overlapping, phase transition at higher temperature stemming from a staging phase with higher $n$. This is discussed further in the {supplementary}. 

The \emph{Bmab} transition is clearly much more gradual (extending more than 100~K) than the fast staging transition for the $x=0.04$ sample. This also seems to apply for the other samples, although the conclusion is less clear because of the fewer datapoints around the transition ($x=0.065$, 0.09) or lack of data above room temperature ($x=0.00$, 0.09). This indicates a different mechanism behind the growth of the \emph{Bmab} volumes than the staging volumes. 
For our range of Sr doping values, both staging and \emph{Bmab} transition temperatures are below the \emph{Bmab} transition temperature of the corresponding oxygen-stoichiometric compound~\cite{Wakimoto2006} (an overview is shown in the {supplementary}).

\begin{figure}[tbp]
  \includegraphics[scale=1]{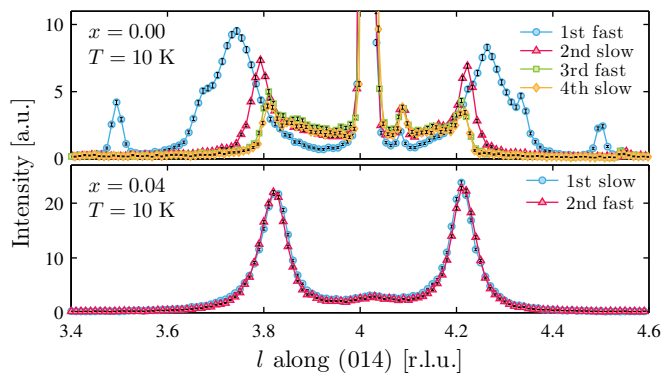}
  \caption{Cooling history dependency of the measured signals for the $x=0.00$ (top) and $x=0.04$ (bottom) samples. 
           The top scattering pattern is changed remarkably from the one shown in the top plot of Fig.~\ref{fig:lowtempoverview} because of the initial fast cool.}
  \label{fig:coolingconditions}
\end{figure}

Finally, an investigation of the effect of cooling rate was performed on the $x=0.00$ and 0.04 samples, as shown in Fig.~\ref{fig:coolingconditions}. The same scans as previously described and shown in Fig.~\ref{fig:lowtempoverview} were performed after fast ($>2$~K/min) and slow ($<2$~K/min) cooling rates. 
We see clear stage 2, 3, and 4 peaks after the 1st fast cool, but also some diffuse signal close to the central \emph{Bmab} peak, while the 2nd slow cool only manages to produce stage 5 on top of the diffuse signal. 
In the sample with Sr, however, the cooling rate has no effect on the scattering pattern, which is dominated by stage 5.


We summarise and discuss our results below. 
The staging numbers found for the oxygen-only doped sample fit well with previously found values for similar samples~\cite{Wells1996}, while the staging observed in the three Sr-doped samples have increasing staging numbers following increasing Sr doping. The peak shoulders observed for the $x=0.09$ sample are analysed as staging, with staging numbers of up to 90, which seem very high. However analysing the widths of the fitted staging peaks (see {supplementary}) for the samples gives an indication of the correlation lengths, which do indeed allow for such long-range interactions. 

The \emph{Bmab} (014) peak observed for the $x=0.04$ sample was heavily suppressed, which could either be due to a Sr doping-induced minimum in the scattering intensity for the reflection or, more likely, be an indication of the staged phase -- which does not exhibit the central \emph{Bmab} peak -- being a very large part of the sample, hence showing only the staging peaks with high intensity~\cite{Wells1996}. 
The \emph{Bmab} peak decreased gradually with temperature similarly to what has been observed previously for optimally Sr-doped La$_{1.85}$Sr$_{0.15}$CuO$_4$~\cite{Braden2001}. The nature of the HTT-LTO transition was described by the same author as largely being displacive and continuous, whereas another author has suggested that substantial CuO$_6$ octahedral tilt (and concommittant charge) disorder is present in optimal and underdoped La$_{2-x}$Sr$_{x}$CuO$_4$~\cite{Bozin1999,Bozin2000}. 
Further studies, possibly including other co-doped samples with low $x$ and a larger \emph{Bmab} volume fraction, are needed in order to rule out either explanation here.
 
An estimate of the relative volume fractions of the staged/non-staged structural phase was given by the ratio of integrated intensities $r_I = I_\text{staging}/I_\text{\emph{Bmab}}$ in Fig.~\ref{fig:lowtempoverview}. For each sample this ratio has a completely different value than the ratio of the superconducting and magnetic volume fractions, which are (in order of low to high $x$) 0.56, 0.79, 4.26, and 1.13 (these values are calculated in the supplementary).  
Thus it is clear that the two types of phase separation -- structural and electronic -- are not directly correlated. 

From the transition temperature of the staging peaks observed in Fig.~\ref{fig:tempdependencies}, it is seen that a higher Sr doping results in a harder binding of the staged phase, observed as a tendency for increasing transition temperature.  This indicates that the Sr could be correlating the staging oxygen, keeping them in a more stable position than what would be the case at La-only sites. Assuming a second order phase transition approach to the transitions for the staging signals, as was argued for the $x=0.04$ sample, 
the critical exponents for the Sr doped samples were all in the range between the 2D and 3D~Ising models. 
At the same time, Sr defects seem to promote faster diffusion of the intercalated oxygen than in oxygen-only doped samples. Thus Sr/O co-doping allows equilibrium to be reached and the same staging pattern evolved even at fast cooling rates, in contrast to oxygen-only doped samples. 
We therefore suggest that future studies of oxygenated samples consider co-doping to avoid temperature hysteresis. 

Possibly our most important observation in the co-doped samples is that the characteristics of the staging patterns, which represent structural modulations along the $c$ axis, do not influence the electronic low-temperature state which is phase-separated between an optimal bulk superconductor with high $T_c\approx 40$~K~\cite{Mohottala2006} and low pinning~\cite{Mohottala2008} and a long-range ordered stripe-like modulated anti-ferromagnet with $T_N \approx T_c \approx 40$~K~\cite{Udby2013PRL}. The co-existence of similar electronic phases have been explained in other cuprates as being intertwined in a spatially self-organised pattern with orderparameters modulated along the $c$ axis~\cite{Fradkin2015,Tranquada2015}. These systems 
however all seem frustrated such that either bulk 3D superconductivity~\cite{Moodenbaugh1988,Tranquada1988} or static anti-ferromagnetic order~\cite{wenPRB85} is suppressed. 
In some cases evidence for 2D superconducting correlations have been observed for temperatures close to our $T_c$~\cite{Li2007,Tranquada2008}.
We infer that intercalation of oxygen apparently lifts both the frustration of the Josephson coupling (since $T_c$ is optimal in all samples) as well as stabilises the magnetic order in a long-range modulated antiferromagnetic pattern with period $\sim$8. 

The influence is however not simple. In a naive picture the staged volume would be oxygen rich (as the superconducting volume with $n_h \sim 0.16$) and the \emph{Bmab} areas hole poor (as the magnetic volume with $n_h \sim 0.125$). We however show in this work that the relative volume fractions of the staged/non-staged phase do not follow the same trend as the relative volume fractions of the magnetic/superconducting domains.


In conclusion, we have shown that our co-doped La$_{2-x}$Sr$_x$CuO$_{4+y}$ samples exhibit staging -- or hints of staging -- with dependence on their Sr content. 
We found that the Sr doping promotes faster oxygen diffusion and stabilises the staged phase, removing temperature hysteresis while at the same time resulting in a tendency for higher transition temperatures for higher doping. 
We also see that the staging number $n$ increases fast with $x$. 
However, the significantly varying characteristics of the staging superstructure which we have discussed in this work, compared with the differently varying characteristics of the electronic phase separation in the same samples, indicate that the interstitial oxygen taking part in the staging superstructure does not (directly) influence the low temperature electronic phase separation. 
Hence, further studies of the intercalated oxygen and the possible connection to electronic phase-separation in similar samples is still needed.

\section*{Acknowledgments} 
\begin{acknowledgments}
  This work was supported by the Danish Agency for Science Technology and Innovation under the Framework Programme on Superconductivity and the Danish Research Council FNU through the instrument center DANSCATT. 
  Work at the University of Connecticut was supported by the U.S. Department of Energy under Contract No. DE-FG02-00ER45801. 
  Samples were grown by F. C. Chou, Center for Condensed Matter Sciences, National Taiwan University, Taipei 10617, Taiwan. 
  The experimental results presented in this work were obtained at the BW5 beamline at the DORIS synchrotron, Deutsches Elektronen-Synchrotron, Hamburg, Germany. 
\end{acknowledgments}

\input{refs_copied_from_bbl.tex} 

\end{document}